\documentclass[sigconf]{acmart}    %
\usepackage{amsmath,amsfonts}

\usepackage{algorithmic}
\usepackage{graphicx}
\usepackage{xspace}
\usepackage{textcomp}
\usepackage{xcolor}
\usepackage{natbib} 
\usepackage{adjustbox}  
\usepackage{algorithm}
\usepackage{algorithmic}
\usepackage{array}
\usepackage{multirow}
\usepackage{booktabs}
\usepackage{color}
\usepackage{subcaption}
\usepackage{threeparttable}
\usepackage[justification=centering]{caption} 

\AtBeginDocument{%
  }
\begin{document}

\def\Figref#1{Figure.~\ref{#1}}
\def\Tabref#1{Table~\ref{#1}}

\newcommand{\YFwrite}[1]{{\color{purple} YF: #1}}
\newcommand{\Wei}[1]{{\color{blue} Wei: #1}}
\newcommand{\yftodo}[1]{{\color{magenta} @YF plz do: #1}}

\newcommand{\ourFramework}{VIS\xspace}
\newcommand{\ours}{BEYOND\xspace}

\def\Eqref#1{Eq.~\eqref{#1}}
\def\Algref#1{Algorithm~\ref{#1}}
\def\Figref#1{Fig.~\ref{#1}}

\newcommand{\var}{{\rm Var}}
\newcommand{\Tr}{^{\rm Tr}}
\newcommand{\vtrans}[2]{{#1}^{(#2)}}
\newcommand{\kron}{\otimes}
\newcommand{\schur}[2]{({#1} | {#2})}
\newcommand{\schurdet}[2]{\left| ({#1} | {#2}) \right|}
\newcommand{\had}{\circ}
\newcommand{\diag}{{\rm diag}}
\newcommand{\invdiag}{\diag^{-1}}
\newcommand{\rank}{{\rm rank}}
\newcommand{\nullsp}{{\rm null}}
\newcommand{\tr}{{\rm tr}}
\renewcommand{\vec}{{\rm vec}}
\newcommand{\vech}{{\rm vech}}
\renewcommand{\det}[1]{\left| #1 \right|}
\newcommand{\pdet}[1]{\left| #1 \right|_{+}}
\newcommand{\pinv}[1]{#1^{+}}
\newcommand{\erf}{{\rm erf}}
\newcommand{\hypergeom}[2]{{}_{#1}F_{#2}}

\renewcommand{\a}{{\bf a}}
\renewcommand{\b}{{\bf b}}
\renewcommand{\c}{{\bf c}}
\renewcommand{\d}{{\rm d}}  
\newcommand{\e}{{\bf e}}
\newcommand{\f}{{\bf f}}
\newcommand{\g}{{\bf g}}
\newcommand{\h}{{\bf h}}
\renewcommand{\k}{{\bf k}}
\newcommand{\m}{{\bf m}}
\newcommand{\n}{{\bf n}}
\renewcommand{\o}{{\bf o}}
\newcommand{\p}{{\bf p}}
\newcommand{\q}{{\bf q}}
\renewcommand{\r}{{\bf r}}
\newcommand{\s}{{\bf s}}
\renewcommand{\t}{{\bf t}}
\renewcommand{\u}{{\bf u}}
\renewcommand{\v}{{\bf v}}
\newcommand{\w}{{\bf w}}
\newcommand{\x}{{\bf x}}
\newcommand{\y}{{\bf y}}
\newcommand{\z}{{\bf z}}
\newcommand{\A}{{\bf A}}
\newcommand{\B}{{\bf B}}
\newcommand{\D}{{\bf D}}
\newcommand{\E}{{\bf E}}
\newcommand{\F}{{\bf F}}
\renewcommand{\H}{{\bf H}}
\newcommand{\I}{{\bf I}}
\newcommand{\J}{{\bf J}}
\newcommand{\K}{{\bf K}}
\renewcommand{\L}{{\bf L}}
\newcommand{\M}{{\bf M}}
\newcommand{\N}{\mathcal{N}}  
\newcommand{\MN}{\mathcal{MN}} 
\newcommand{\Acal}{\mathcal{A}}
\newcommand{\Ocal}{\mathcal{O}}
\newcommand{\Dcal}{\mathcal{D}}
\newcommand{\Ycal}{\mathcal{Y}}
\newcommand{\Zcal}{\mathcal{Z}}
\newcommand{\Fcal}{\mathcal{F}}
\newcommand{\Vcal}{\mathcal{V}}
\newcommand{\Lcal}{\mathcal{L}}
\newcommand{\Tcal}{\mathcal{T}}
\newcommand{\Gcal}{\mathcal{G}}
\newcommand{\Hcal}{\mathcal{H}}
\newcommand{\Scal}{\mathcal{S}}
\newcommand{\Xcal}{\mathcal{X}}

\renewcommand{\O}{{\bf O}}
\renewcommand{\P}{{\bf P}}
\newcommand{\Q}{{\bf Q}}
\newcommand{\R}{{\bf R}}
\renewcommand{\S}{{\bf S}}
\newcommand{\T}{{\bf T}}
\newcommand{\V}{{\bf V}}
\newcommand{\W}{{\bf W}}
\newcommand{\X}{{\bf X}}
\newcommand{\Y}{{\bf Y}}
\newcommand{\Z}{{\bf Z}}
\newcommand{\Mcal}{{\mathcal{M}}}
\newcommand{\Wcal}{{\mathcal{W}}}
\newcommand{\Ucal}{{\mathcal{U}}}

\newcommand{\bfLambda}{\boldsymbol{\Lambda}}

\newcommand{\bsigma}{\boldsymbol{\sigma}}
\newcommand{\balpha}{\boldsymbol{\alpha}}
\newcommand{\bpsi}{\boldsymbol{\psi}}
\newcommand{\bphi}{\boldsymbol{\phi}}
\newcommand{\boldeta}{\boldsymbol{\eta}}
\newcommand{\Beta}{\boldsymbol{\eta}}
\newcommand{\btau}{\boldsymbol{\tau}}
\newcommand{\bvarphi}{\boldsymbol{\varphi}}
\newcommand{\bzeta}{\boldsymbol{\zeta}}
\newcommand{\bepsilon}{\boldsymbol{\epsilon}}

\newcommand{\blambda}{\boldsymbol{\lambda}}
\newcommand{\bLambda}{\mathbf{\Lambda}}
\newcommand{\bOmega}{\mathbf{\Omega}}
\newcommand{\bomega}{\mathbf{\omega}}
\newcommand{\bPi}{\mathbf{\Pi}}

\newcommand{\btheta}{\boldsymbol{\theta}}
\newcommand{\bpi}{\boldsymbol{\pi}}
\newcommand{\bxi}{\boldsymbol{\xi}}
\newcommand{\bSigma}{\boldsymbol{\Sigma}}

\newcommand{\bgamma}{\boldsymbol{\gamma}}
\newcommand{\bGamma}{\mathbf{\Gamma}}

\newcommand{\bmu}{\boldsymbol{\mu}}
\newcommand{\1}{{\bf 1}}
\newcommand{\0}{{\bf 0}}

\newcommand{\bs}{\backslash}

 \newcommand{\notS}{{\backslash S}}
 \newcommand{\nots}{{\backslash s}}
 \newcommand{\noti}{{\backslash i}}
 \newcommand{\notj}{{\backslash j}}
 \newcommand{\nott}{\backslash t}
 \newcommand{\notone}{{\backslash 1}}
 \newcommand{\nottp}{\backslash t+1}

\newcommand{\notk}{{^{\backslash k}}}
\newcommand{\notij}{{^{\backslash i,j}}}
\newcommand{\notg}{{^{\backslash g}}}
\newcommand{\wnoti}{{_{\w}^{\backslash i}}}
\newcommand{\wnotg}{{_{\w}^{\backslash g}}}
\newcommand{\vnotij}{{_{\v}^{\backslash i,j}}}
\newcommand{\vnotg}{{_{\v}^{\backslash g}}}
\newcommand{\half}{\frac{1}{2}}
\newcommand{\msgb}{m_{t \leftarrow t+1}}
\newcommand{\msgf}{m_{t \rightarrow t+1}}
\newcommand{\msgfp}{m_{t-1 \rightarrow t}}

\newcommand{\proj}[1]{{\rm proj}\negmedspace\left[#1\right]}
\newcommand{\argmin}{\operatornamewithlimits{argmin}}
\newcommand{\argmax}{\operatornamewithlimits{argmax}}

\newcommand{\dif}{\dfrac{\dfrac{\mathrm}{den}}{den}{d}}
\newcommand{\abs}[1]{\lvert#1\rvert}
\newcommand{\norm}[1]{\lVert#1\rVert}

\newcommand{\ie}{{i.e.,}\xspace}
\newcommand{\eg}{{e.g.,}\xspace}
\newcommand{\etc}{{etc.}\xspace}

\newcommand{\EE}{\mathbb{E}}
\newcommand{\dr}[1]{\nabla #1}
\newcommand{\VV}{\mathbb{V}}
\newcommand{\sbr}[1]{\left[#1\right]}
\newcommand{\rbr}[1]{\left(#1\right)}
\newcommand{\cmt}[1]{}

\newcommand{\bi}{{\bf i}}
\newcommand{\bj}{{\bf j}}
\newcommand{\bK}{{\bf K}}
\newcommand{\Vtr}{\mathrm{Vec}}

\newcommand{\cov}{{\rm Cov}}	

\newtheorem{Proposition}{Proposition}
\newtheorem{Lemma}{Lemma}
\newtheorem{Corollary}{Corollary}
\newtheorem{Remark}{Remark}
\newtheorem{Assumption}{Assumption}
\newtheorem{Property}{Property}

\newcommand{\RR}{\mathbb{R}}
\newcommand{\KL}{\mathrm{KL}}

\newcommand{\bPsi}{\boldsymbol{\Psi}}
\newcommand{\bXi}{\boldsymbol{\Xi}}
\newcommand{\btx}{\textbf{\textit{x}}}
\newcommand{\bty}{\textbf{\textit{y}}}
\newcommand{\btz}{\textbf{\textit{z}}}
\newcommand{\btk}{\textbf{\textit{k}}}

\newcommand{\bupsilon}{\boldsymbol{\upsilon}}

\newcommand{\GP}{\mathcal{GP}}
\newcommand{\TGP}{\mathcal{TGP}}
\newcommand{\TNcal}{\mathcal{TN}}

\title{The Name of the Title Is Hope}
\title{Beyond the Yield Barrier: A Variational Framework For Optimal Importance Sampling Yield Estimation }
\title{Beyond the Yield Barrier: Variational Analysis For Optimal Importance Sampling Yield Estimation }
\title{Beyond the Yield Barrier: Optimal Importance Sampling In Yield Analysis }
\title{Beyond the Yield Barrier: Variational Analysis For Optimal Importance-Sampling-Based Yield Analysis }
\title{Beyond the Yield Barrier: Variational Importance Sampling Yield Analysis\footnotemark[2]}
\renewcommand{\thefootnote}{\fnsymbol{footnote}}

\settopmatter{printacmref=false} 

\author{Yanfang Liu$^{1}$, Lei He$^{2}$, Wei W. Xing$^{3}$}
\authornote{Corresponding author.}
\affiliation{
\institution{
$^{1}$ School of Integrated Circuit Science and Engineering, Beihang University \city{Beijing} \country{China}\\
$^{2}$ Eastern Institute of Technology \city{Ningbo} \country{China}, $^{3}$ SoMas, The University of Sheffield \country{U.K.}
}
}
\email{liuyanfang@buaa.edu.cn, 
lhe@eitech.edu.cn,
w.xing@sheffield.ac.uk}

\begin{abstract}
Optimal mean shift vector (OMSV)-based importance sampling methods have long been prevalent in yield estimation and optimization as an industry standard.
However, most OMSV-based methods are designed heuristically without a rigorous understanding of their limitations.
To this end, we propose \ourFramework, the first variational analysis framework for yield problems, enabling a systematic refinement for OMSV. For instance, \ourFramework reveals that the classic OMSV is suboptimal, and the optimal/true OMSV should always stay beyond the failure boundary, which enables a free improvement for all OMSV-based methods immediately.
Using \ourFramework, we show a progressive refinement for the classic OMSV including incorporation of full covariance in closed form, adjusting for asymmetric failure distributions, and capturing multiple failure regions, each of which contributes to a progressive improvement of more than 2$\times$. 
Inheriting the simplicity of OMSV, the proposed method retains simplicity and robustness yet achieves up to 29.03$\times$ speedup over the state-of-the-art (SOTA) methods. We also demonstrate how the SOTA yield optimization, ASAIS, can immediately benefit from our True OMSV, delivering a 1.20$\times$ and 1.27$\times$ improvement in performance and efficiency, respectively, without additional computational overhead.
\end{abstract}
\keywords{Yield Estimation, Importance Sampling, Variational Analysis}
\maketitle

\section{Introduction}

\footnotetext[2]{The title of this paper is a homage to the seminal 2008 work of Lara Dolecek et al., ``Breaking the simulation barrier: Sram evaluation through norm minimization'' which laid the groundwork for contemporary yield analysis. 
The key word Beyond is twofold: 1) the optimal shift vector literately should lie beyond the failure boundary and 2) the performance of the proposed method can go beyond the classic one.}

{
With the continual advancement of integrated circuit technology, microelectronic devices are shrinking to submicrometer scales. This trend has elevated the significance of random process variations, including intra-die mismatches, doping fluctuations, and threshold voltage variations, as critical factors in circuit design.
In modern circuit designs, particularly in scenarios like SRAM cell arrays where certain cells can be replicated millions of times, addressing yield concerns has become paramount. Efficient yield estimation methods are crucial for providing accurate and rapid failure rate assessments in the presence of specific process variations.

Monte Carlo (MC) simulation, the industry-standard baseline, is commonly employed for yield estimation. MC involves running SPICE (Simulation Program with Integrated Circuit Emphasis) simulations with parameters drawn from the process variation distribution millions of times, counting failures to obtain precise estimates. However, MC is computationally intensive and becomes impractical for practical problems where the yield can be as low as $10^{-5}$, a common setup in a 65nm SRAM cell array.

One pivotal avenue toward efficient yield estimation involves harnessing the potential of machine learning (ML) to construct data-driven surrogate models, approximating the unknown indicator function. Active learning techniques are then employed to iteratively refine the surrogate. 
Notably, \cite{AYEBO} leverages a Gaussian process (GP) for modeling the underlying performance functions and employs an entropy reduction strategy in active learning. 
Absolute shrinkage deep kernel learning (ASDK) replaces the GP with a nonlinear-correlated deep kernel method and feature selection to identify crucial features for focused analysis~\cite{ASDK}. 
\cite{LRTA} adopts a low-rank tensor approximation (LRTA) to approximate the performance function. 
Recently, Optimal Manifold Importance Sampling (OPTIMIS) proposes to use normalizing flow model to capture the optimal failure manifold~\cite{nf}.
Despite their success, surrogate-based methods are less favored due to their susceptibility to instability and the demand for data for surrogate model training. 
Surrogate-based methods are vulnerable to the highly nonlinear optimization problems inherent in model training, which, if not addressed correctly, can yield erroneous surrogate models and consequently inaccurate yield estimation scenarios the industry cannot afford. 

Currently, the most extensively applied methods in the industrial landscape for Electronic Design Automation (EDA) tools are the Scaled-sigma Sampling (SSS) and Optimal Mean Shift Vector (OMSV)-based techniques due to their simplicity and robustness ~\cite{EDAtool}.
SSS generates random samples from a distorted distribution for which the standard deviation is scaled up to reduce the samples of simulations and enhance estimation efficiency \cite{SSS}.
OMSV-based methods employ the importance sampling (IS) technique, constructing a Gaussian distribution with OMSV as the mean and using it as the proposal distribution from which samples are drawn to accelerate yield estimation. 
For the OMSV-based methods, finding OMSV is most critical.
Minimum Norm Importance Sampling (MNIS) identifies the Minimum Norm (MN) failure sample, a.k.a. the most probable failure point (MPFP), as the OMSV~\cite{MNIS}. 
Owing to its success, many subsequent studies redirect their focus towards finding MN-OMSV.
Gradient Importance Sampling (GIS) enhances efficiency in finding MN-OMSV by employing gradient descent~\cite{GIS}.
Fast Sensitivity Importance Sampling (FSIS) uses transient sensitivity analysis instead of gradient-descent to find MN-OMSV~\cite{ASAIS}.
To resolve the challenge of multiple failure regions, 
Hyperspherical Clustering and Sampling (HSCS) employs clustering to identify them and finds the MN-OMSV for each failure region~\cite{HSCS}.
To keep MN-OMSV updated with more simulations, Adaptive Importance Sampling (AIS) introduces a dynamically updated sampling distribution, enhancing the accuracy of yield estimation~\cite{AIS}.
By integrating the key ideas of HSCS and AIS, Adaptive Clustering and Sampling (ACS) further enhances the efficiency of yield estimation with multiple failure regions~\cite{ACS}.
The importance of OMSV-based yield estimation is self-evident and it is \textit{de facto} an industry standard due to its simplicity and robustness, which also lays the foundation for the state-of-the-art (SOTA) yield optimization, \eg All Sensitivity Adversarial
Importance Sampling (ASAIS)~\cite{ASAIS}.

Despite their success, most progresses are heuristic in nature, it is unclear when or where their assumptions are valid. To this end, we introduce the first variational analysis framework for yield analysis: Variational Importance Sampling (\ourFramework), which serves as an unifying framework for various SOTA methods.

Based on \ourFramework, we discover a surprising fact: the true/optimal OMSV is not the widely used MN-OMSV first proposed in MNIS in 2008~\cite{MNIS}, but rather, it always stays beyond the failure boundary (vs. on the failure boundary as MN-OMSV).
This insight instantly grants us free improvement without extra costs for SOTA methods built on the OMSV assumption, \eg ASAIS.
\ourFramework further reveals that MNIS and SSS are special cases of the same assumed proposal distribution, and there exists a closed-form solution where these two methods can be unified.
To showcase the power of \ourFramework, we further introduce extra refinements, including a skew normal distribution to further boost efficiency and a mixture of distributions to handle multiple failure regions. 
%
%
In summary, the novelty of this work includes:
\begin{enumerate}
  \item \ourFramework, the first variational analysis framework for IS-based yield analysis, paving the way for the rigorous design and analysis for computational yield problems, with the following novelty as demonstration.
  \item True OMSV, the calibrated version of the canonical MN-OMSV, generating a free-lunch speedup up to 10$\times$.
  \item Full SSS, a complete version of SSS, which admits a closed-form solution for the covariance matrix, offering another up to 2$\times$ speedup at no extra computational cost.  
  \item Skew Normal (SN) OMSV, introducing asymmetric distribution to offer an extra 1.4$\times$ speedup.
  \item Mixture of Skew Normals (MSN) OMSV, which is used to handle multiple failure region challenges.
  {\item The combination of (2)-(5) as a novel yield estimation method, which we call \ours (to suggest the importance of True OMSV and superior performance).}
  \item Variational-ASAIS, demonstrating how \ourFramework can immediately improve SOTA yield optimization, ASAIS, by 1.20$\times$ in performance and 1.27$\times$ in efficiency.
  \item The superiority of \ours is validated on multiple SRAM and analog circuits with thoughtful experiments, ablation study and robustness study, which demonstrate a 2.50$\times$-29.03$\times$ speedup (9.78$\times$ on average) and a 0.11\%-24.49\% improvement (7.33\% on average) in yield estimation accuracy. 
\end{enumerate}

}

\section{Background}
\subsection{Problem Definition}
\cmt
{Denote $\x=[x^{(1)},x^{(2)},\cdots,x^{(D)}]^T \in \Xcal$ as the variation process parameter, and $\Xcal$ the variation parameter space. 
$\Xcal$ is generally a high-dimensional space (\ie large $D$);
Each variable in $\x$ denotes the variation parameters of a circuit during manufacturing, \eg length or width of PMOS and NMOS transistors. 
In general, $\x$ are considered mutually independent Gaussian distributed, 
\begin{equation}
     p(\x) = (2\pi)^{\frac{D}{2}}\exp \left(- ||\x||^2 /2 \right).
\end{equation}
Given a specific value of $\x$, we can evaluate the circuit performance $\y$ (\eg  memory read/write time and amplifier gain) through SPICE simulation, $ \y=\f(\x)$,
where $\f(\cdot)$ is the SPICE simulator, which is considered an expensive and time-consuming black-box function;
$\y=[y^{(1)},y^{(2)},\cdots,y^{(K)}]^T$ are the collections of circuit performance based on the simulations.
When $K$ metrics are all smaller than or equal to their respective thresholds (predefined by designers) $\boldsymbol{t}$, \ie $y^{(k)} \leq t^{(k)}$ for $k=1,\cdots,K$, the circuit is considered as a successful design; otherwise, it is a failure one.
We use failure indicator $I(\x)$, which is 1 if $\x$ leads to a failure design and 0 otherwise, to denote the failure status of a circuit.
Finally, the ground-truth failure rate $\hat{P}_f $ is:
\begin{equation}
    \hat{P}_f = \int_\mathcal{X} I\left( \x \right) p(\x) d\x.
\end{equation}}

Let $\x=[x^{(1)},x^{(2)},\cdots,x^{(D)}]^T \in \Xcal$ denote the variation variables, with $\Xcal$ representing the parameter space for such variations. Typically, $\Xcal$ is a high-dimensional space, denoted as $D$, where each element within the vector $\x$ signifies a specific manufacturing-related parameter affecting a circuit, such as the dimensions (length or width) of PMOS and NMOS transistors.
In the context of our analysis, we make a general assumption that the elements of $\x$ are statistically independent and follow a Gaussian distribution:
\begin{equation}
    p(\x) = (2\pi)^{\frac{D}{2}}\exp \left(- \frac{1}{2}{||\x||^2} \right).
\end{equation}

Given $\x$, we can measure the performance of the circuit, denoted as $\y$ (e.g., metrics like memory read/write time and amplifier gain), by using SPICE simulation.
Denote this as $\y = \f(\x)$, where $\f(\cdot)$ represents the SPICE simulator.
If $\y$ satisfies all pre-defined conditions $\boldsymbol{t}$, \eg $y^{(k)} \leq t^{(k)}$ for $k=1,\cdots,K$, then the design is considered a success; otherwise, it is a failure.
Introducing an indication function $I(\x)$ to denote the failure case, the ground-truth failure rate $\hat{P}_f $ is defined as:
\begin{equation}
\hat{P}_f = \int_\mathcal{X} I\left( \x \right) p(\x) d\x.
\end{equation}


\subsection{Monte Carlo Yield Estimation}

The direct calculation of the yield is intractable due to the unknown $I(\x)$.
A straightforward approach to estimate the failure rate is MC, which involves sampling $\x_i$ from $p(\x)$ and evaluating the failure rate by the ratio of failure: 
\begin{equation}
    P_f = \frac{1}{N} \sum_{i=1}^N I(\x_i),
\end{equation}
where $\x_i$ is the $i$-th sample from $p(\x)$, and $N$ is the number of samples.
To obtain an estimate of $1- \varepsilon$ accuracy with $1-\delta$ confidence, $N \approx { \log(1/\delta)}/{\varepsilon^2 \hat{P}_f }$ is required.
For a modest $90\%$ accuracy $(\varepsilon=0.1)$ with $90\%$ confidence $(\delta=0.1)$, we need $N \approx 100/\hat{P}_f$ samples, which is infeasible in practice for small $\hat{P}_f$, say, $10^{-5}$.
We can also see this intuitively from the fact that it requires on average $1/\hat{P_f}$ samples just to observe a failure event.

\subsection{Importance Sampling Yield Estimation}
{
In contrast to sampling directly from the distribution $p(\mathbf{x})$, IS-based methods utilize a proposal distribution $q(\mathbf{x})$ to draw samples and estimate the failure rate as follows:
\begin{equation}
  \label{eq:IS}
  \begin{aligned}
    P_f  &= \int_\mathcal{X} \frac{I(\x) p(\x)} { q(\x) }  q(\x) d\x 
    \approx \frac{1}{N} \sum_{i=1}^N \frac{I(\x_i) p(\x_i)} {q(\x_i)} = \sum_{i=1}^N {I(\x_i) w(\x_i)},
  \end{aligned}
\end{equation}
where $\x_i$ are samples drawn from $q(\x)$ and are used to approximate the integral as in MC. For convenience, we define the importance weight $w(\x)=p(\x) / q(\x)$.
\Eqref{eq:IS} is proved to be more efficient than traditional MC, provided that the proposal distribution $q(\mathbf{x})$ is thoughtfully selected.
}

\subsection{MN-OMSV}

\begin{figure*}[t]
  \centering
  \begin{subfigure}{0.24\textwidth}  
  \subcaption{MN-OMSV}
  \label{fig:MNIS}
  \includegraphics[width=1\linewidth]{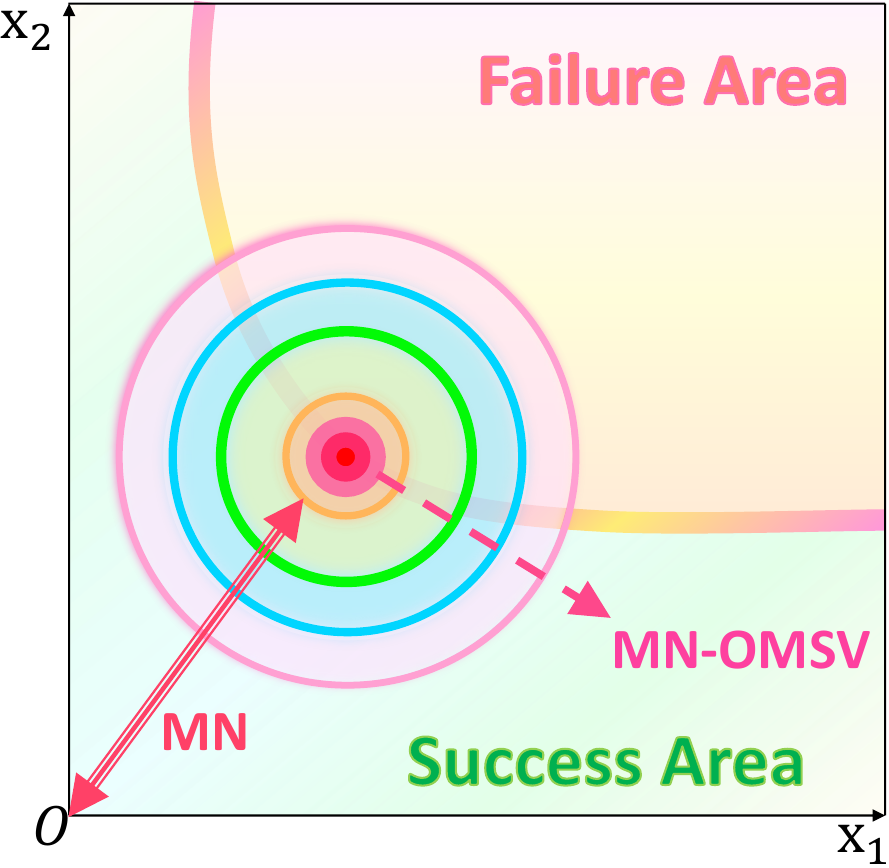}
  \end{subfigure}
  \begin{subfigure}{0.24\textwidth}
    \subcaption{True OMSV}
    \label{fig:to}
    \includegraphics[width=1\linewidth]{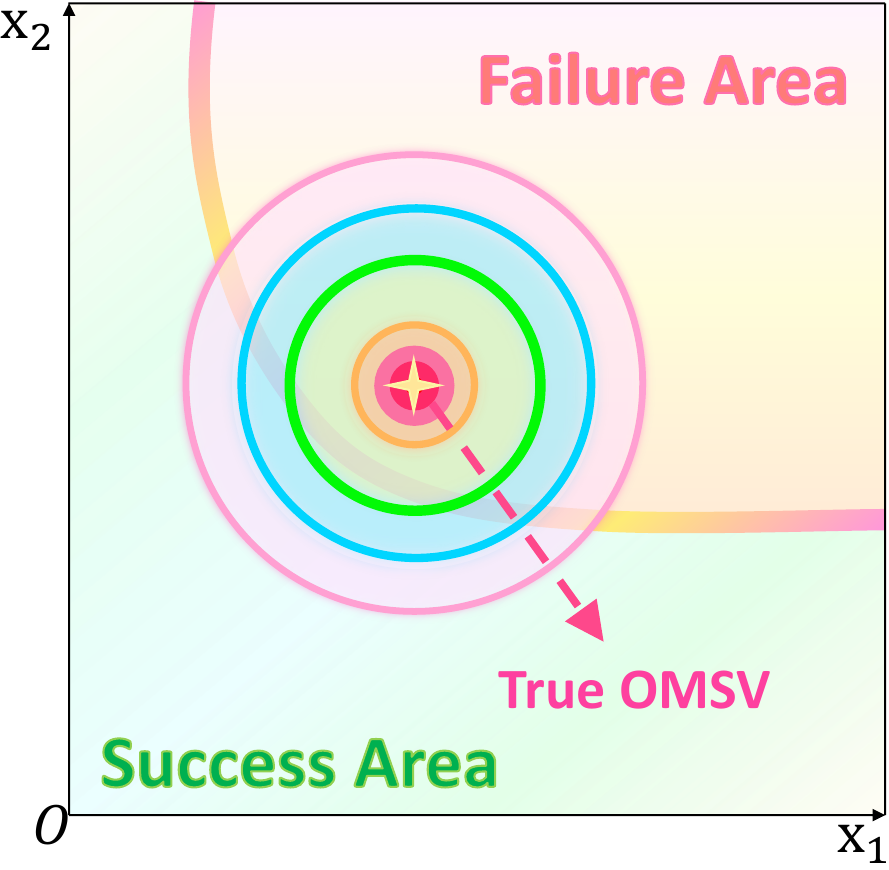}
    \end{subfigure}
  \begin{subfigure}{0.24\textwidth}
    \subcaption{True OMSV+Full SSS}
    \label{fig:tos}
    \includegraphics[width=1\linewidth]{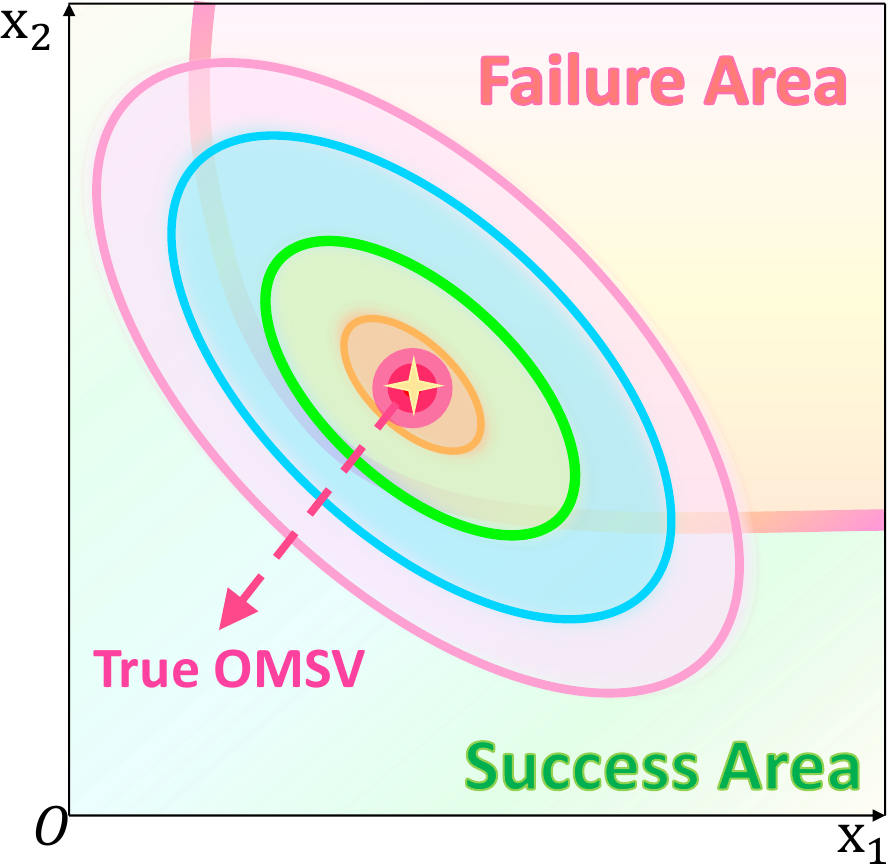}
    \end{subfigure}
    \begin{subfigure}{0.24\textwidth}
      \subcaption{True OMSV+Full SSS+SN}
      \label{fig:toss}
    \includegraphics[width=1\linewidth]{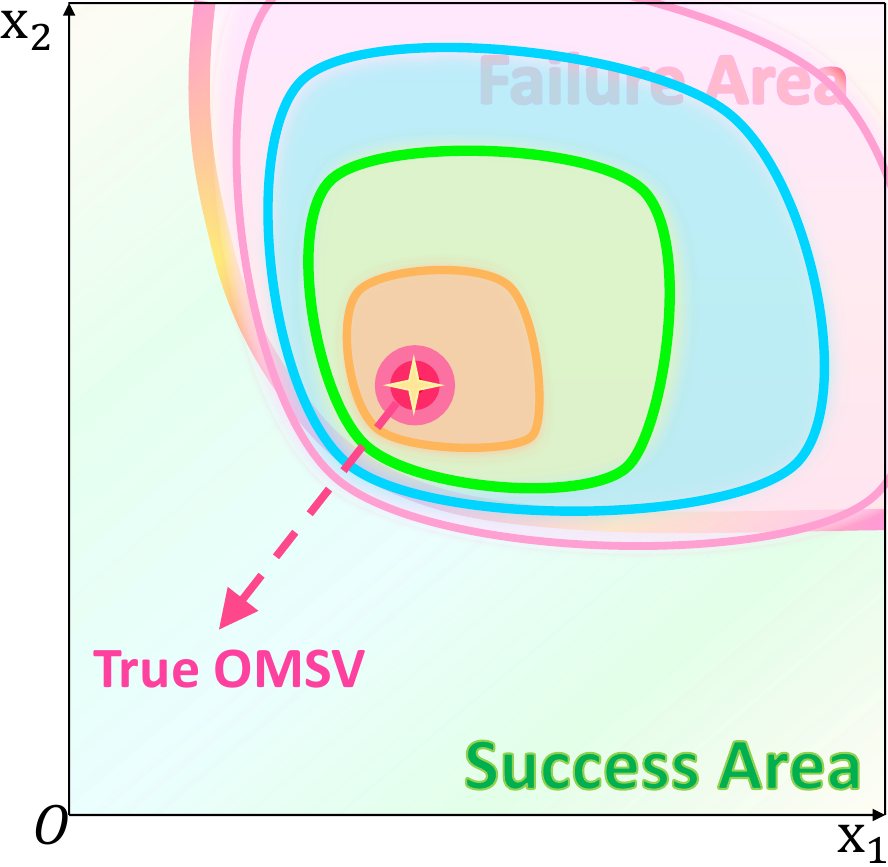}
    \end{subfigure}\\
    \vspace{-0.15in}
    \caption{Illustration of progressive refinement of the classic MN-OMSV using \ourFramework. 
    }
    \label{fig:illu}
    \vspace{-0.15in}
\end{figure*}

A canonical approach to design a proposal distribution \( q(\mathbf{x}) \) involves employing a normal distribution \( \mathcal{N}(\boldsymbol{\mu}, \I) \), effectively shifting the original Gaussian distribution centered at the origin to \( \boldsymbol{\mu} \). The optimal shift vector \( \boldsymbol{\mu}^* \), referred to as OMSV (a.k.a. MPFP), can be computed by solving the following optimization problem as delineated in MNIS \cite{MNIS}:
\begin{equation}
  \label{eq:OSV}
  \boldsymbol{\mu}^* = \arg \min ||\mathbf{x}||^2  \quad \text{s.t.} \quad I(\mathbf{x}) = 1,
\end{equation}
where \( ||\mathbf{x}||^2 = \sum_{d=1}^{D} (x^{(d)})^2 \) represents the Euclidean norm. As illustrated in \Figref{fig:MNIS}, MNIS essentially uses the existing failure samples with the minimal norm as the OMSV to propose new samples.

\section{Proposed Approach}

Despite that MN-OMSV is intuitive, it was never rigorously justified in the literature.
We first introduce a variational analysis framework for yield analysis, \ourFramework, which will provide an analysis of the MN-OMSV.
We then use \ourFramework to progressively improve OMSV with rigorous mathematical analysis, which is also illustrated in \Figref{fig:illu}.

\subsection{Variational Analysis of IS Yield Estimation}

From \Eqref{eq:IS}, we can see that the optimal proposal distribution $q^*(\x)$ is the one that minimizes the approximate variance, \ie
\begin{equation}
  q^*(\x)= \argmin_{q} \EE_{q} \left[ w^2(\x) \left( I(\x)-\hat{P}_f \right)^2 \right].
\end{equation}
Utilizing the Lagrange multiplier rule for the calculus of variations, we can show that the optimal proposal distribution is given by
\begin{equation}
  \label{eq:optimal_proposal} 
  q^*(\x) = p(\x)I(\x)/\hat{P}_f.
\end{equation}
Thus, the optimal IS yield estimation is equivalent to minimizing the Kullback-Leibler (KL) divergence between the true optimal proposal distribution, $q^*(\mathbf{x})$, and its approximate counterpart, $q(\mathbf{x})$
\begin{equation}
  \label{eq:KL}
  \KL(q^*(\x)||q(\x)) =  \EE_{q^*(\x)} \left[ \log q^*(\x) \right] - \EE_{q^*(\x)} \left[ \log q(\x) \right],
\end{equation}
with a limited number of samples.
Although alternative divergence metrics (\eg \( \KL(q(\x)||q^*(\x)) \) and Wasserstein distance) exist, this work specifically employs \( \KL(q^*(\x)||q(\x)) \). 
As we will see later, this choice yields closed-form solutions to avoid extensive hyperparameter tuning and computational overhead, a common issue with modern ML-based approaches. 
Notably, $\mathbb{E}_{q^*(\mathbf{x})} \left[ \log q^*(\mathbf{x}) \right]$ is an unknown constant denoting the entropy of the optimal proposal distribution, leaving the optimization focus on the second term.
Thus, minimization of the KL divergence is equivalent to maximizing $\EE_{q^*(\x)} \left[ \log q(\x) \right]$,
which admits an approximated solution by only keeping the failure samples
\begin{equation}
  \label{eq:beyond}
  \begin{aligned}
 \int  p(\x)I(\x)/\hat{P}_f   \log \left( q(\x)  \right) \d \x 
 \approx \frac{1}{\hat{P}_f} \sum_{i=1}^{N'} g(\x_i) p(\x_i) \log \left( q(\x_i) \right)
  \end{aligned}
\end{equation}
where $\x_i$ are failure samples, \ie $I(\x_i)=1$, $N'$ is the number of failure samples, and $g(\x_i)$ is distribution that generates the samples.
To better explore the failure regions, which are crucial for approximating the integral with a small number of samples, this work uses a uniform distribution, \ie $g(\x_i)=1$.

\Eqref{eq:beyond} is the key insight of this work---to provide a variational framework, VIS, using a numerical approximation to the ideal KL divergence. No assumption of the unknown function $I(\mathbf{x})$ is made and the approximation is exact when the number of samples $N'$ approaches infinity.

\subsection{True OMSV}

Based on \ourFramework, let's now revisit the OMSV \cite{MNIS} and assume that the proposal distribution is a Gaussian with a mean shift $\bmu$, \ie
$q(\x) = \mathcal{N}(\x|\bmu, \I).$
Substituting this $q(\x)$ into \Eqref{eq:beyond} and taking the derivative w.r.t $\bmu$, we achieve the optimal $\bmu$
\begin{equation}
\label{eq:optimal mu}
    \bmu = \frac{\sum_{i=1}^{N'} p(\x_i) \x_i}{\sum_{i=1}^{N'} p(\x_i)}.
\end{equation}

This elegant closed-form solution reveals that the ``True OMSV'' that maximizes the objective function is the weighted average of the failure samples and \textbf{it always resides beyond the failure boundary NOT on the failure boundary}.
We can also see that the importance of each failure sample decreases as it moves away from the origin, explaining why the classic MN-OMSV can still work well as a special case of using just one failure sample with the maximum weight.

\subsection{Full SSS}

With \ourFramework, it is now possible to transcend the limitations of a fixed variance Gaussian distribution for the proposal distribution. The concept of employing varying variances for the proposal is not novel itself, as it has been previously explored in the pioneer work \cite{SSS}.
However, this method only considers a single variance scaling factor for all dimensions, thereby ignoring the correlations between dimensions and leading to suboptimal performance.
Moreover, the selection of variance relies on a heuristic approach and expert knowledge, which is not practical for deployment in real-world applications.
Here, we take a more ambitious step by assuming a full covariance matrix for the proposal distribution, \ie
$q(\x) = \mathcal{N}(\x|\bmu, \bSigma),$
which may seem overkill and can lead to overfitting. As we will see soon, the covariance matrix $\bSigma$ will admit a closed-form solution under \ourFramework.
Substituting the proposal into \Eqref{eq:beyond}, taking the derivative w.r.t $\bmu$ and $\bSigma$ and setting them to zero, we can derive the optimal $\bmu$ and $\bSigma$.
Not surprisingly, the optimal $\bmu$ is exactly \Eqref{eq:optimal mu}, and the optimal $\bSigma$ is
\begin{equation}
  \label{eq:optimal S}
  \begin{aligned}
    \mathbf{\Sigma} = \frac{\sum_{i=1}^{N'} p(\x_i) (\x_i - \bmu)(\x_i - \bmu)^T}{\sum_{i=1}^{N'} p(\x_i)}.
  \end{aligned}
\end{equation}
As a special case of the full covariance matrix, we can also derive the optimal variance for SSS by forcing $\bSigma=\sigma^2\I$ and get the optimal
\begin{equation}
    \sigma^2 = \frac{\sum_{i=1}^{N'} p(\x_i)(\x_i - \bmu)^T (\x_i - \bmu)}{ \sum_{i=1}^{N'} p(\x_i)}.
\end{equation}
A diagonal form of $\bSigma$ can also be assumed. Since the full covariance matrix $\bSigma$ is available in closed form, there is no need to use a diagonal form unless overfitting becomes an issue, which was not encountered in our experiments.

We can see how powerful \ourFramework is,
as it allows us to derive closed-form solutions for the optimal mean and covariance of the proposal distribution. In other words, we can derive better solutions than the conventional methods, while preserving tractability to minimize the computational costs and model complexity, which are important key merits the industry is looking for.

\subsection{Skew Normal Distribution}
\label{sec:skew_normal}
A ubiquitous assumption made in the OMSV-based literature is that the proposal distribution is a symmetric Gaussian distribution.
While this assumption is convenient for the analysis and simple enough to prevent overfitting, it can also lead to a significant reduction in efficiency by proposing many samples in the failure region (about 50\%, see \Figref{fig:tos} for an example).
This is intuitive to see because at least half of the samples will be generated inside the failure region, for the simplest cases with one failure region. In practice, it can get worse when the failure regions have a narrow shape. 
This issue is not resolved in the literature due to the lack of proper analysis tools and the difficulty in deriving a feasible solution.
With \ourFramework, we can now take a further step by amending the proposal distribution to be the multivariate skew normal distribution, which is a generalization of the normal distribution and can better fit the optimal proposal distribution {(see \Figref{fig:toss})}. The Probability Density Function (PDF) of the multivariate skew normal is
\begin{equation}
  \label{eq:skew_normal}
  \mathcal{SN}(\mathbf{x} | \boldsymbol{\mu}, \boldsymbol{\Sigma}, \boldsymbol{\alpha}) = 2 \phi(\mathbf{x}; \boldsymbol{\mu}, \boldsymbol{\Sigma}) \Phi(\boldsymbol{\alpha}^T \mathbf{x}).
\end{equation}
Here:
\( \phi(\mathbf{x}; \boldsymbol{\mu}, \boldsymbol{\Sigma}) \) is the PDF of the normal distribution with mean vector \( \boldsymbol{\mu} \) and covariance matrix \( \boldsymbol{\Sigma} \);
\( \Phi(\cdot) \) is the cumulative distribution function (CDF) of the standard normal distribution;
\( \boldsymbol{\alpha} \) is a \( D \)-dimensional vector of shape parameters. The vector \( \boldsymbol{\alpha} \) determines the skewness in each dimension. When \( \boldsymbol{\alpha} = \mathbf{0} \), the multivariate skew normal distribution reduces to the standard multivariate normal distribution. 

To get the parameters of the multivariate skew normal distribution, we can substitute \Eqref{eq:skew_normal} into \Eqref{eq:beyond}, take the derivative w.r.t $\bmu$, $\bSigma$ and $\boldsymbol{\alpha}$ and set them to zero.
\begin{equation}
  \label{eq:optimal sn}
  \begin{aligned}
  \argmax_{\bmu,\boldsymbol{\Sigma}, \boldsymbol{\alpha} } \sum_{i=1}^{N'} p(\x_i)   \log \left(\mathcal{SN}(\mathbf{x}_i | \boldsymbol{\mu}, \boldsymbol{\Sigma}, \boldsymbol{\alpha}) \right).
  \end{aligned}
\end{equation}
Unfortunately, this does not lead to a closed-form solution for the parameters, which is not surprising as the estimation of the parameters in the multivariate skew normal distribution itself is a known challenge.
To deliver a practical solution, we use the mean and covariance estimated from the previous sections and only optimize \Eqref{eq:optimal sn} w.r.t the shape parameter $\boldsymbol{\alpha}$ using gradient descent.
This turns out to be an excellent workaround as it fits well with our motivation to have an asymmetric proposal distribution, instead of deriving a well-fitting distribution from scratch.

\subsection{Mixture of Skew Normal Distribution}
Finally, all simple OMSV-based methods can only deal with a single failure region, which poses a significant limitation for real-world applications.
This problem can be simply resolved by introducing a mixture of skew-normal distribution, \ie
\begin{equation}
  \label{eq: mix normal}
  q(\x) = \sum_{m=1}^M w_m \mathcal{SN}(\mathbf{x} | \boldsymbol{\mu}_m, \boldsymbol{\Sigma}_m, \boldsymbol{\alpha}_m),
\end{equation}
where $w_i$ is the weight and $M$ is the number of mixture components. Substituting \Eqref{eq: mix normal} into \Eqref{eq:beyond} and doing the optimization, we can derive the optimal mixture of skew normal distribution. However, this optimization is extremely difficult.
As a workaround, we first cluster the failure samples using silhouette coefficient \cite{sc_cluster,cl2}, which automatically determines the number of clusters and the cluster label for each failure sample. The weight $w_m$ is approximated by the number of samples in each cluster divided by the total number of failure samples. Finally, the parameters $\{\boldsymbol{\mu}_m, \boldsymbol{\Sigma}_m, \boldsymbol{\alpha}_m\}$ for each cluster are optimized by \Eqref{eq:optimal sn}, \Eqref{eq:optimal S} and \Eqref{eq:optimal mu}.

\begin{algorithm}[t]
  \caption{\ours Algorithm}
  \begin{algorithmic}[1] \label{algo:all}
      \REQUIRE SPICE-based indication function $I(\x)$
      \STATE Use Onion Sampling \cite{nf} to form initial  failure samples set $\Dcal$
      \REPEAT
          \STATE Update iteration $t=t+1$
          \STATE Use silhouette coefficient to get $M$ clusters and weight $w_m$ 
          \STATE Fit each cluster with a skew normal distribution using \Eqref{eq:optimal sn}, \Eqref{eq:optimal S} and \Eqref{eq:optimal mu} and form $q(\x)$ with \Eqref{eq: mix normal}
          \STATE Draw $K$ samples from ${{q}^{t}(\x)}$ and calculate importance weights: $w_k^t = {I(\x_k)p(\x_k)}/{q^t(\x_k)}$ for $k=1,2,\ldots,K$.
          \STATE Estimate failure rate $\hat{P}_{f}=\frac{1}{tK}\sum\limits_{j=1}^{t}{\sum\limits_{k=1}^{K}{{w}^{t}_{k}}}$.
          \STATE Update failure sample collection $\Dcal$
      \UNTIL{Figure of Merit (FOM), ${std(\hat{P}_{f})}/{\hat{P}_{f}}<0.1$}
      \RETURN{Failure rate estimation $\hat{P}_{f}$}
  \end{algorithmic}
\end{algorithm}

\subsection{Complexity and Implementation}
Given $N$ as the number of failure samples, the computation of the silhouette coefficient is $\Ocal(ND)$.
The computation of True OMSV and Full SSS is $\Ocal(N')$ and $\Ocal(N' D^2)$, respectively, where $N'$ is the number of failure samples in a cluster. Updating the skew normal shape parameters is $\Ocal(N')$ each iteration.
The overall algorithm is summarized in \Algref{algo:all}. Note that the algorithm is flexible by using only the True OMSV, Full SSS.

\subsection{Calibration of SOTA Yield Optimization}
Many SOTA yield optimization methods rely on yield estimation by MN-OMSV~\cite{AOSM,ASAIS,MNFV}, which has been revealed to be sub-optimal in this work. Nonetheless, just by using the True OMSV instead of the MN-OMSV and keeping other parts of the method unchanged, we can achieve better performance for no extra cost.
We choose the latest advanced OMSV-based yield optimization method, ASAIS~\cite{ASAIS}, as the baseline method. It optimizes the design parameter $\z$ by maximizing the following objective function:
\begin{equation}
    \argmax_\z ||\bmu (\z)||^2,
\end{equation}
where $\bmu (\z)$ is the OMSV computed by \Eqref{eq:OSV} for design parameter $\z$. This optimization is solved by gradient descent with gradient
$2\bmu\frac{\partial \bmu}{\partial \z}$ given by adjoint method implemented in the SPICE solver.
According to the True OMSV, we simply modify the computation of $\bmu (\z)$ using \Eqref{eq:optimal mu}. The gradient is then given by the weighted sum of the gradient of all failure samples. We call the modified method Variational-ASAIS.

\def\Figref#1{Fig.~\ref{#1}}
\def\Tabref#1{Table~\ref{#1}}

\section{Experimental Results}
In this section, we conduct a comprehensive evaluation of the accuracy and efficiency of our method, namely \ours, in yield estimation on three benchmark circuits: a 6T-SRAM, an operational transconductance amplifier (OTA) and a 6-bit 6T-SRAM array circuit.
To ensure a SOTA comparison, we implement seven SOTA methods as comparative baselines: MNIS~\cite{MNIS}, HSCS~\cite{HSCS}, AIS~\cite{AIS}, ACS~\cite{ACS}, LRTA~\cite{LRTA}, ASDK~\cite{ASDK}, and OPTIMIS~\cite{nf}. MC serves as the gold standard for estimating the true failure rate. We also utilize the Figure of Merit (FoM), denoted as $\rho$, calculated as $\rho = {\mathrm{std}(P_f)}/{P_f}$, where $\mathrm{std}(P_f)$ is the standard deviation of the estimated failure rate. 
FoM serves as the termination criterion for all methods and we use $\rho=0.1$ following~\cite{MNIS,AOSM,HSCS}. 
For the assessment, speedup is computed as $\frac{\# Sim_{MC}}{\# Sim}$, and the relative error rate is $({P}_f-\hat{P}_{f_{MC}}) / \hat{P}_{f_{MC}}$.

In all of our experiments, we conduct ten random seed experiments for each method (ensuring the same seeds for all methods). The final failure rate estimation is obtained by taking the average across these ten experiments. Additionally, we select the best-performing result from the ten random experiments for each method and use it to create a visualization of the iterative estimation of failure rate and its FoM.
We implement the baselines with their default configurations, and where necessary, we fine-tune hyperparameters to optimize performance.
All experiments are conducted on a Windows system with an AMD 7950x CPU and 32GB RAM.

  \begin{figure}[t]
  \centering
  \includegraphics[width=0.9\linewidth]{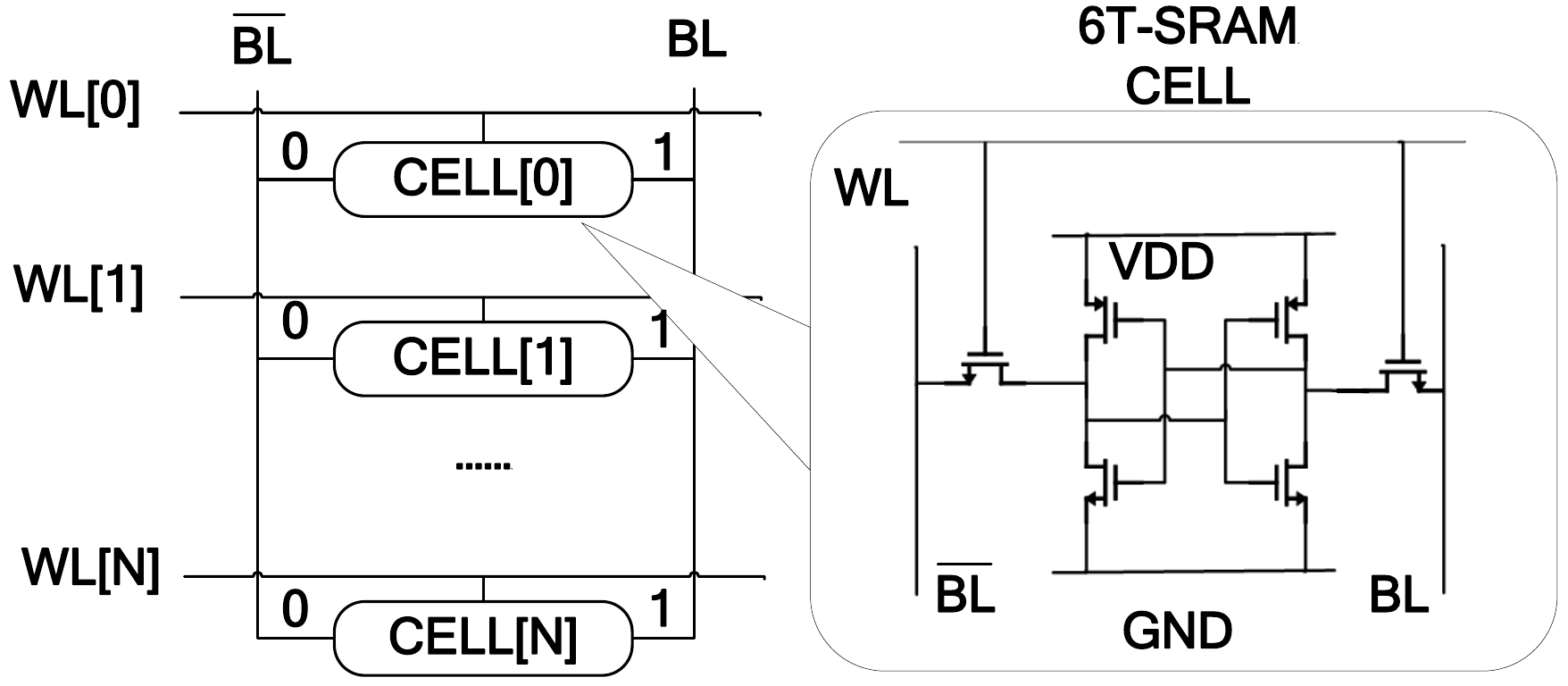}
  \vspace*{-0.15in}
  \caption{The structure of SRAM column circuit}
  \label{fig:6t}
\end{figure}

\begin{figure}[t]
  \centering
  \includegraphics[width=1.\linewidth]{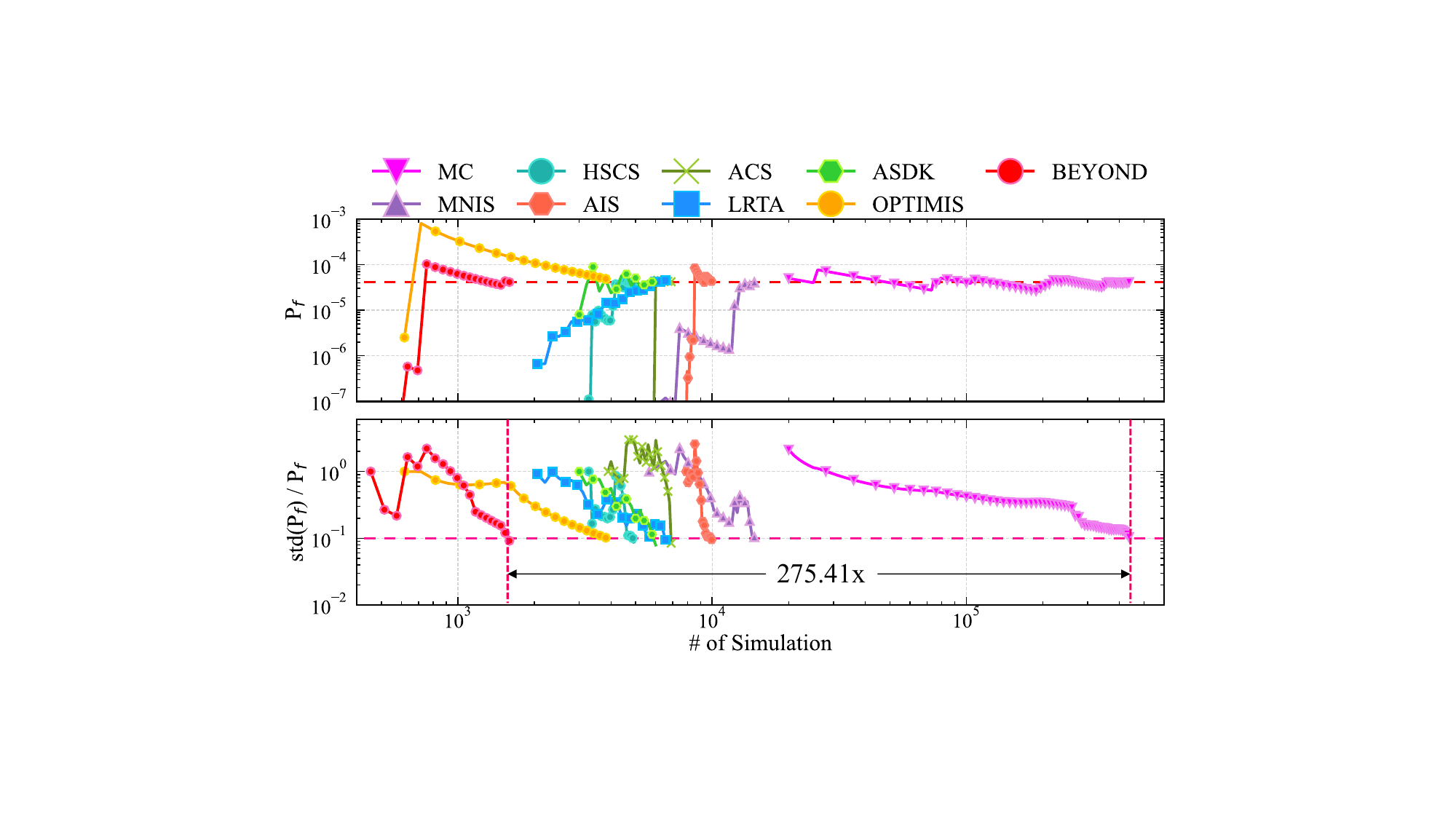}
  \vspace*{-0.2in}
  \caption{Failure rate estimation with FoM on 6T-SRAM}
  \label{fig:18dim}
  \vspace*{-0.1in}
\end{figure}

\begin{table}[t]
  \centering
  \caption{Yield Estimation Results on 6T-SRAM}
  \label{tab:18dim}
    \vspace{-0.15in}
    \begin{adjustbox}{width=1.\columnwidth,center}
    \renewcommand\arraystretch{1}
    \setlength{\tabcolsep}{3.5pt}
  \begin{tabular}{l|cccc}
  \toprule[1.5pt]
  \midrule[0.5pt]
  \multicolumn{1}{l|}{Model} & Fail. Rate & Rel. Err. & \# Sim & Speedup \\ 
  \midrule[1pt]
MC & 4.99e-5 & - & 406240 & 1$\times$ \\
MNIS & 4.81e-5 & 3.61\% & 10030 & 40.50$\times$ \\
HSCS & 4.86e-5 & 2.61\% & 4152 & 97.84$\times$ \\
AIS & 4.85e-5 & 2.81\% & 9702 & 41.87$\times$ \\
ACS & 4.70e-5 & 5.81\% & 9620 & 42.23$\times$ \\
LRTA & 4.86e-5 & 2.61\% & 6130 & 66.27$\times$ \\
ASDK & 4.85e-5 & 2.81\% & 6640 & 61.18$\times$ \\
OPTIMIS & 4.93e-5 & 1.18\% & 3916 & 103.74$\times$ \\
\ours & \textbf{4.98e-5} & \textbf{0.16\%} & \textbf{1564} & \textbf{259.74$\times$}\\
\midrule[0.5pt]
  \bottomrule[1.5pt]
  \end{tabular}
  \end{adjustbox}
  \vspace{-0.1in}
  \end{table}

\subsection{6T-SRAM Circuit} 

The 6T-SRAM bit cell, illustrated in \Figref{fig:6t}, is implemented in a 45nm CMOS process, which includes six transistors. Each transistor has three independent random variables: threshold voltage, mobility, and gate oxide thickness, which are critically impactful on yield among all variation parameters. As a result, the circuit encompasses 18 independent random variables. In our experiments, we focus on the delay time of SRAM read/write as the performance metric of interest.

The yield estimation experimental results are shown in \Tabref{tab:18dim}, and the evolution of failure rate convergence and FoM evaluation is depicted in \Figref{fig:18dim}.
As shown in \Tabref{tab:18dim}, it is evident that \ours achieves the most accurate estimation with minimal simulations. In terms of accuracy, \ours exhibits a relative error rate as low as 0.16\%, improving the accuracy by  1.02\%-5.65\% against other baselines. In terms of efficiency, \ours achieves a speedup of up to 259.74$\times$ compared to MC, and demonstrates a speedup of 2.50$\times$-6.41$\times$ compared to other baselines.

\begin{figure}[t]
  \centering
  \includegraphics[width=1.\linewidth]{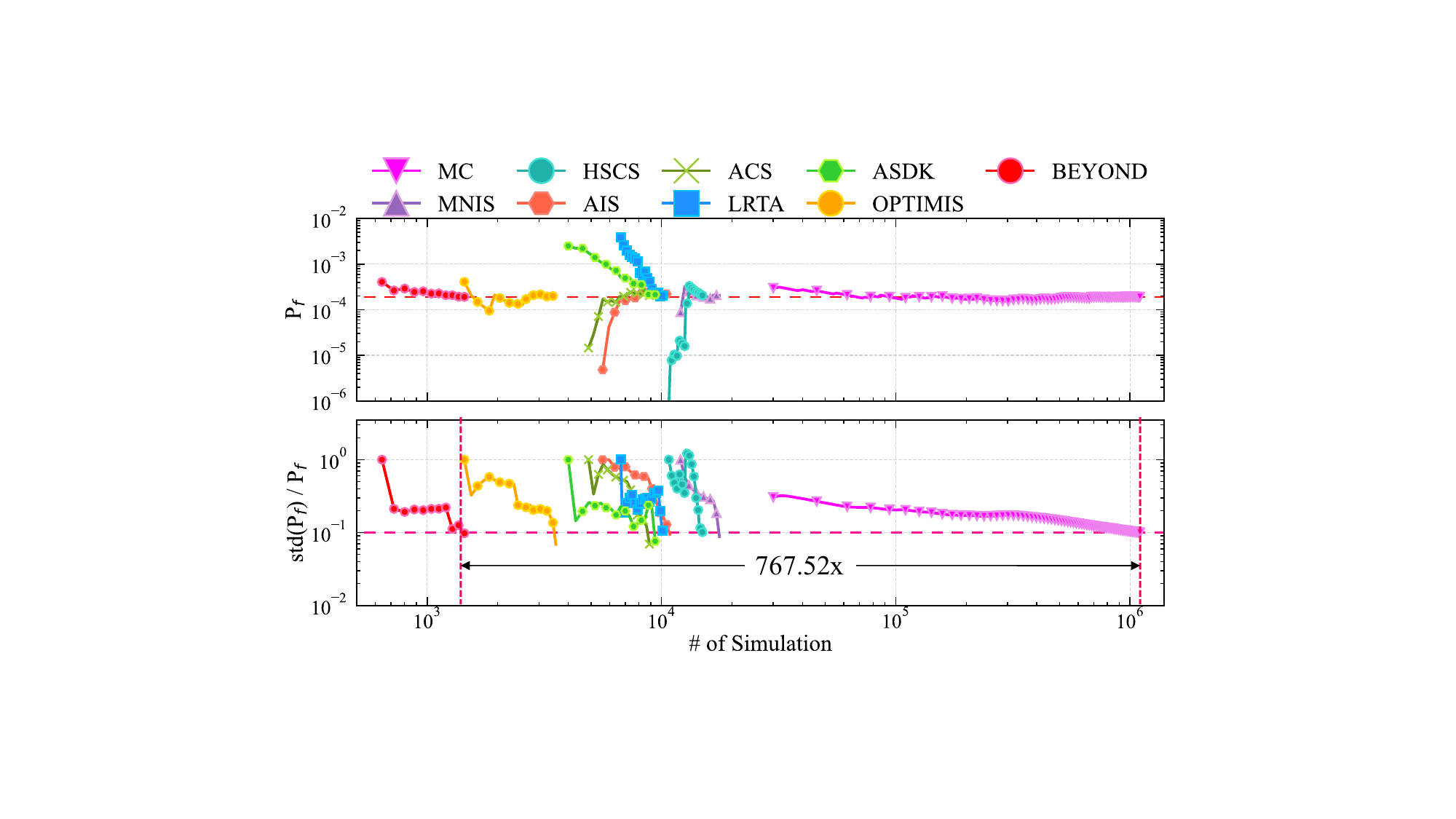}
  \vspace*{-0.2in}
  \caption{Failure rate estimation with FoM on OTA}
  \label{fig:56dim}
  \vspace*{-0.1in}
\end{figure}

\begin{table}[t]
  \centering
  \caption{Yield Estimation Results on OTA}
  \label{tab:56dim}
    \vspace{-0.15in}
    \begin{adjustbox}{width=1\columnwidth,center}
    \renewcommand\arraystretch{1}
    \setlength{\tabcolsep}{3.5pt}
  \begin{tabular}{l|cccc}
  \toprule[1.5pt]
  \midrule[0.5pt]
  \multicolumn{1}{l|}{Model} & Fail. Rate & Rel. Err. & \# Sim & Speedup \\ 
  \midrule[1pt]
MC & 1.89e-4 & - & 1102000 & 1$\times$ \\
MNIS & 1.64e-4 & 11.94\% & 21065 & 52.31$\times$ \\
HSCS & 1.70e-4 & 10.15\% & 17950 & 61.39$\times$ \\
AIS & 1.74e-4 & 8.18\% & 11178 & 98.59$\times$ \\
ACS & 1.78e-4 & 5.59\% & 11053 & 99.70$\times$ \\
LRTA & 2.04e-4 & 7.94\% & 10100 & 109.11$\times$ \\
ASDK & 2.14e-4 & 11.68\% & 9600 & 114.79$\times$ \\
OPTIMIS & 1.92e-4 & 1.57\% & 4126 & 267.09$\times$ \\
\ours & \textbf{1.90e-4} & \textbf{0.41\%} & \textbf{1441} & \textbf{764.75$\times$}\\
\midrule[0.5pt]
  \bottomrule[1.5pt]
  \end{tabular}
  \end{adjustbox}
  \vspace{-0.05in}
  \end{table}

\begin{figure}[h]
  \centering
  \includegraphics[width=0.9\linewidth]{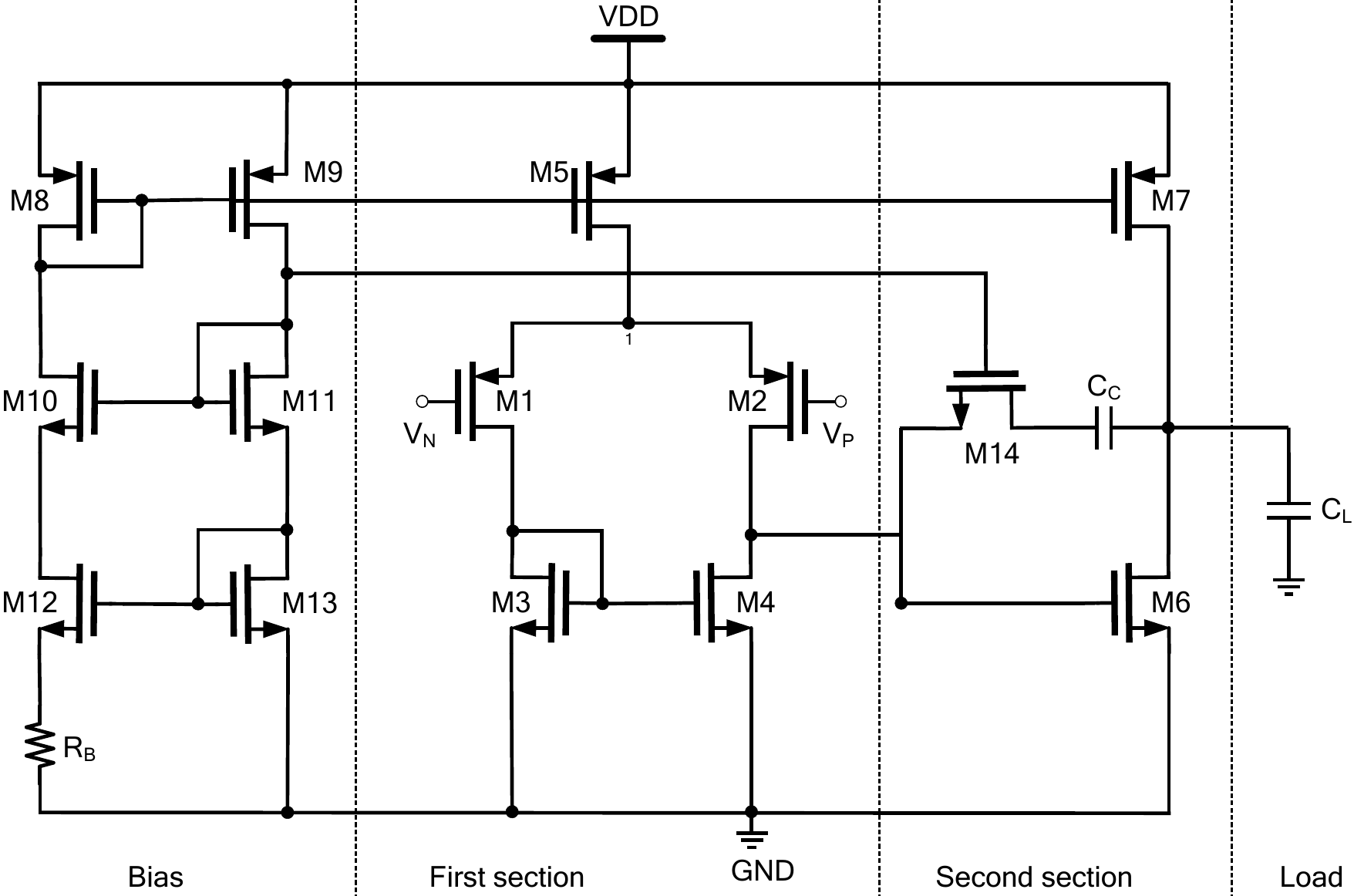}
  \vspace*{-0.1in}
  \caption{Operational Transconductance Amplifier Circuit}
  \label{fig:ota}
  \vspace*{-0.15in}
\end{figure}

\subsection{Operational Transconductance Amplifier}

The OTA circuit, depicted in the \Figref{fig:ota}, contains 14 transistors. Each transistor has four process variation parameters: oxide thickness, threshold voltage, and deviations in length and width due to process variations. Consequently, this circuit comprises 56 independent random variables. In our experiments, the performance of interest is the quiescent current $I_Q$ at $27^{\circ}C$. The yield estimation results are detailed in \Tabref{tab:56dim}, and the evolution of failure rate convergence and FoM evaluation is illustrated in \Figref{fig:56dim}.

The results indicate that \ours consistently delivers highly accurate estimation with the reduced simulations in the analog circuit. In accuracy terms, \ours achieves a relative error rate as low as 0.41\%, enhancing the accuracy by 1.16\%-11.53\% over other baselines. In efficiency terms, \ours realizes a speedup of up to 764.75$\times$ relative to MC and shows a speedup of 2.86$\times$-14.62$\times$ compared to other baselines. These results underscore the robustness of \ours in varied circuit complexities.

\begin{figure}[t]
  \centering
  \includegraphics[width=1.\linewidth]{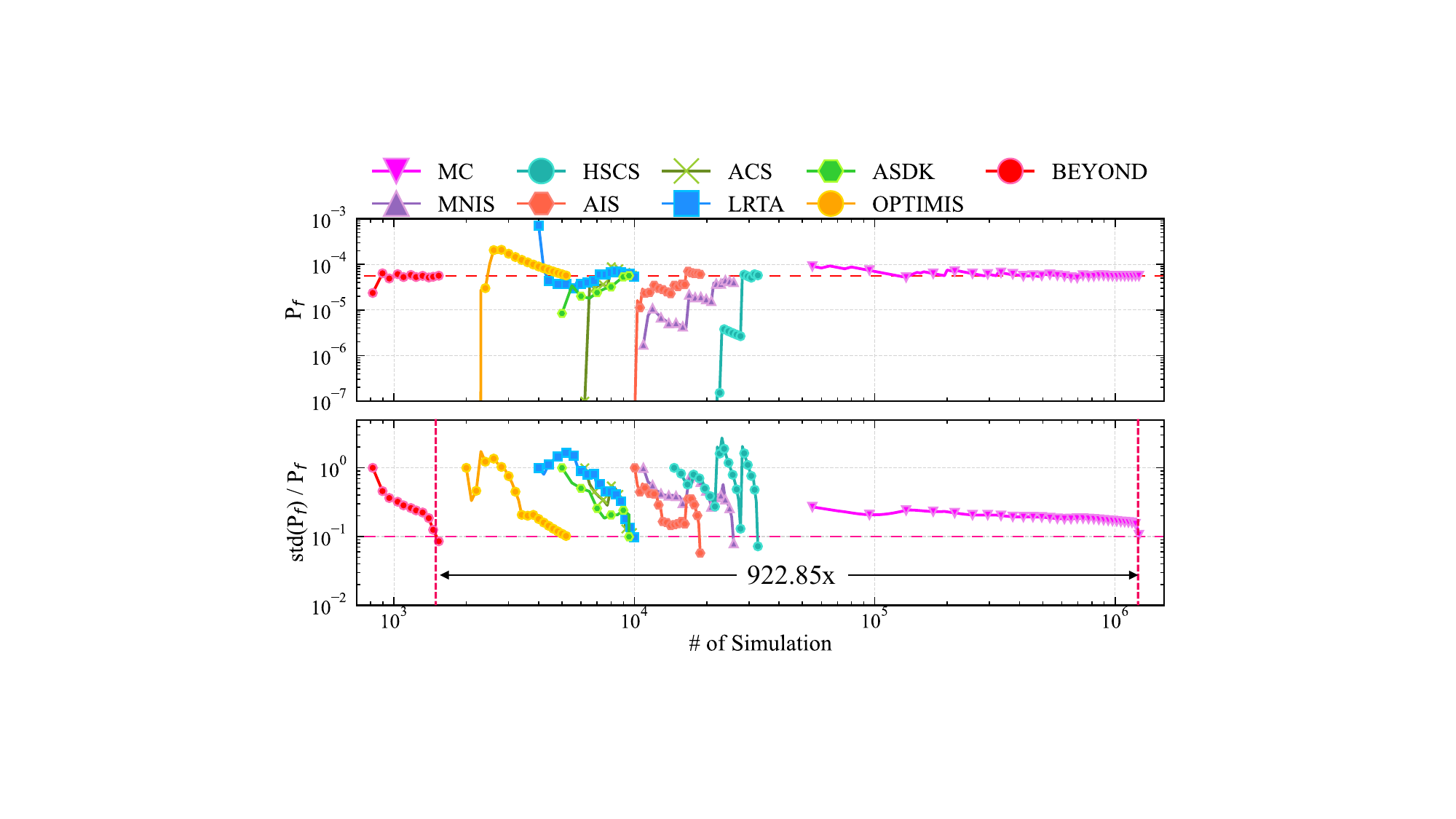}
  \vspace*{-0.2in}
  \caption{Failure rate estimation with FoM on 6-bit array}
  \label{fig:108dim}
  \vspace*{-0.1in}
\end{figure}

\begin{table}[t]
  \centering
  \caption{Yield Estimation Results on 6-bit 6T-SRAM Array}
  \label{tab:108dim}
    \vspace{-0.15in}
    \begin{adjustbox}{width=1\columnwidth,center}
    \renewcommand\arraystretch{1}
    \setlength{\tabcolsep}{3.5pt}
  \begin{tabular}{l|cccc}
  \toprule[1.5pt]
  \midrule[0.5pt]
  \multicolumn{1}{l|}{Model} & Fail. Rate & Rel. Err. & \# Sim & Speedup \\ 
  \midrule[1pt]
MC & 5.62e-5 & - & 1417500 & 1$\times$ \\
MNIS & 4.94e-5 & 12.03\% & 45174 & 31.38$\times$ \\
HSCS & 4.21e-5 & 25.09\% & 47090 & 30.10$\times$ \\
AIS & 4.37e-5 & 22.19\% & 15996 & 88.62$\times$ \\
ACS & 4.92e-5 & 12.44\% & 14060 & 100.82$\times$ \\
LRTA & 5.96e-5 & 6.05\% & 12300 & 115.24$\times$ \\
ASDK & 5.87e-5 & 4.44\% & 12500 & 113.40$\times$ \\
OPTIMIS & 5.66e-5 & 0.71\% & 5300 & 267.45$\times$ \\
\ours & \textbf{5.59e-5} & \textbf{0.60\%} & \textbf{1622} & \textbf{873.92$\times$}\\
\midrule[0.5pt]
  \bottomrule[1.5pt]
  \end{tabular}
  \end{adjustbox}
  \vspace{-0.15in}
  \end{table}


\begin{table*}[t]
\centering
\caption{Yield Optimization Comparison Results on Adder Circuit}
\vspace*{-0.15in}
\label{tab:opti}
\begin{adjustbox}{width=2.12\columnwidth,center}
  \renewcommand\arraystretch{1.3}
  \setlength{\tabcolsep}{2.4pt}
\begin{tabular}{l!{\vline width 0.8pt}ccccccc !{\vline width 0.8pt}ccccccc}
\toprule[1.5pt]
\midrule[0.5pt]
Case & \multicolumn{7}{c!{\vline width 0.8pt}}{1 (Higher Specification)} & \multicolumn{7}{c}{2 (Lower Specification)} \\ 
\hline
Metric & \multicolumn{4}{c|}{Failure Rate} & \multicolumn{3}{c!{\vline width 0.8pt}}{\# Simulation} & \multicolumn{4}{c|}{Failure Rate} & \multicolumn{3}{c}{\# Simulation} \\ \hline
Method & Best & Worst & Mean & \multicolumn{1}{c|}{Std} & Best & Worst & Mean & Best & Worst & Mean & \multicolumn{1}{c|}{Std} & Best & Worst & Mean \\ \midrule[1pt]

WEIBO & 7.50e-7 & 2.08e-5 & 6.80e-6 & \multicolumn{1}{c|}{8.61e-6} & 2121 & 4861 & 3626 & 1.50e-5 & 1.90e-5 & 1.74e-5  & \multicolumn{1}{c|}{1.36e-6} & 2681 & 4391 & 3536 \\

MESBO &  5.00e-8 & 1.50e-7 & 6.00e-8  & \multicolumn{1}{c|}{ 3.00e-8} &4070  & 11200 & 8640 &  1.03e-5 & 2.00e-5 & 1.70e-5  & \multicolumn{1}{c|}{2.83e-6} &  4220 & 7870 & 5687  \\

KDEBO & 5.00e-8 & 1.30e-6 & 3.45e-7   & \multicolumn{1}{c|}{4.54e-7} & 10000 &10000  & 10000 &  1.10e-5 & 1.46e-4 & 5.24e-5  & \multicolumn{1}{c|}{5.40e-5} &  8000 & 8000 & 8000  \\

BYA &  4.00e-8 & \textbf{5.00e-8} & 4.50e-8  & \multicolumn{1}{c|}{\textbf{5.00e-9}} &11000  &11000  & 11000 & 1.70e-5 & 1.75e-5 & 1.72e-5 & \multicolumn{1}{c|}{\textbf{2.29e-7}} & 8000 & 8000 & 8000 \\

ASAIS &  \textbf{2.50e-8} & 7.50e-8 & 4.75e-8  & \multicolumn{1}{c|}{2.08e-8} & 2417 &  2427& 2422 &  9.00e-6 & 2.60e-5 & 1.83e-5  & \multicolumn{1}{c|}{5.88e-6} &  2412 & 2420 & 2415  \\

\textbf{V.-ASAIS} &  \textbf{2.50e-8} & \textbf{5.00e-8} & \textbf{4.00e-8}  & \multicolumn{1}{c|}{1.22e-8} & \textbf{1875} & \textbf{1987} &  \textbf{1912}& \textbf{7.00e-6} & \textbf{1.20e-5} & \textbf{8.50e-6} & \multicolumn{1}{c|}{1.50e-6} & \textbf{1968} &\textbf{ 1992 }& \textbf{1981} \\ 
\midrule[0.5pt]
\bottomrule[1.5pt]
\end{tabular}
\end{adjustbox}
\vspace{-0.1in}
\end{table*}

\subsection{6-bit 6T-SRAM Array Circuit}

Building upon the \ours validated in the 6T-SRAM bit cell experiments, we expand to a higher complexity with the 6-bit 6T-SRAM array circuit, which has six such bit cells. This circuit contains a total of 108 variational parameters, offering a comprehensive view that incorporates peripheral circuit influences to enhance failure rate estimation accuracy. Results are captured in \Tabref{tab:108dim}, and the evolution of failure rate convergence and FoM evaluation is shown in \Figref{fig:108dim}.

For the higher-dimensional circuit, \ours still maintains its leading edge. Accuracy-wise, \ours achieves a relative error rate as low as 0.16\% and an accuracy enhancement of 0.11\%-24.49\% over the baselines. Efficiency-wise, \ours exhibits a remarkable speedup, reaching up to 873.92$\times$ compared to MC and achieving a speedup of 3.27$\times$-29.03$\times$ compared to other baselines.
These results not only reinforce the precision of \ours but also underscore its efficiency in managing the heightened complexity of advanced SRAM architectures.

\begin{figure}[t]
  \centering
  \includegraphics[width=1.05\linewidth]{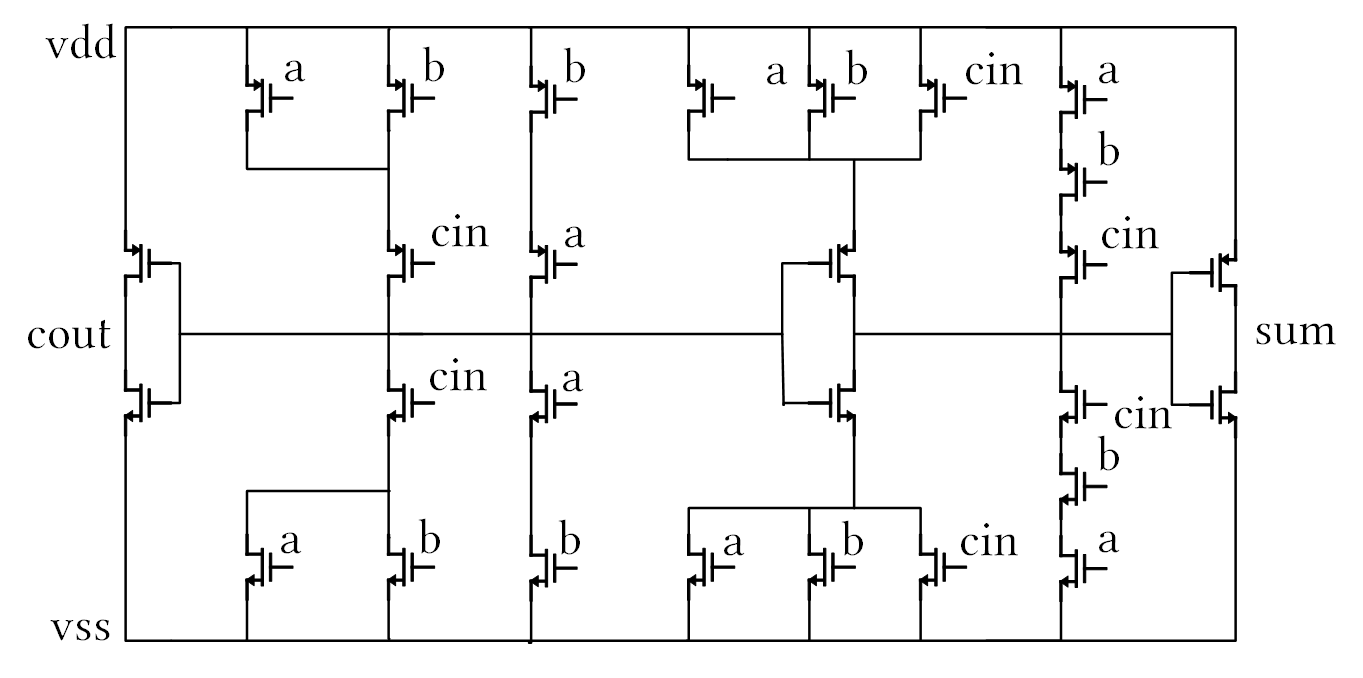}
  \vspace*{-0.3in}
  \caption{The structure of Adder circuit}
  \label{fig:adder}
  \vspace*{-0.2in}
\end{figure}
\subsection{Yield Optimization on Adder Circuit}

Building upon our successful yield estimation experiments, we now focus on yield optimization through the ASAIS optimization flow~\cite{ASAIS}, herein referred to as Variational-ASAIS. We benchmark Variational-ASAIS against five SOTA yield optimization methods: Weighted Expected Improvement Bayesian Optimization (WEIBO)~\cite{EIBO}, Max-value Entropy Search Bayesian Optimization (MESBO)~\cite{MESBO}, Kernel Density Estimator Bayesian Optimization (KDEBO)~\cite{KDEBO}, Bayesian Yield Analysis (BYA)~\cite{AYEBO}, and ASAIS for a comprehensive comparison.

We conduct the yield optimization experiments on an adder circuit, illustrated in \Figref{fig:adder}.
The adder circuit
comprises 28 MOS transistors, each subject to three variational parameters, totaling 84 variables. Our design parameters focus on the width and length of these transistors. 
We assess the yield by examining the time-to-threshold (TT) performance, which involves simulating the transient response until the sum output attains a specified threshold voltage.
To validate the optimization performance of each method, we conduct experiments with ten different random seeds to reduce random fluctuations (ensuring the same seeds for all methods). 
Furthermore, we employ two distinct circuit specifications - a higher case (Case 1) and a lower case (Case 2) - to assess the robustness of all methods.
The optimal design is validated using 4e7 and 1e6 MC simulations  for Case 1 and Case 2, respectively. The results are summarized in \Tabref{tab:opti}.

For Case 1 on the high specification, BYA achieves the lowest standard deviation and its worst-case result equating to that achieved by Variational-ASAIS, which consistently leads in performance. 
While ASAIS also posts competitive optimization results, Variational-ASAIS outstrips all baselines when considering mean performance, boasting a 1.13$\times$-170$\times$ improvement over the other baselines. 
In efficiency terms, Variational-ASAIS proves to be the most resource-sparing, surpassing the baselines by 1.27$\times$ to 5.75$\times$.
For Case 2 on the lower specification, BYA continues to show a good result in the lowest standard deviation. But in other aspects, all baselines are inferior to Variational-ASAIS, which achieves a 2$\times$-6.16$\times$ optimization performance improvement with a 1.22$\times$-4.03$\times$ speedup compared to other baselines based on the mean results.
Collectively, these findings highlight Variational-ASAIS's exceptional optimization prowess while conserving simulations.

\begin{table}[h]
  \centering
  \caption{Comparison of Computational Time (CPU Hours) }
  \label{tab:com_time}
  \vspace{-0.15in}
  \begin{adjustbox}{width=1\columnwidth,center}
    \renewcommand\arraystretch{1.2}
    \setlength{\tabcolsep}{2.5pt}
  \begin{tabular}{l|cccccccc}
  \toprule[1.5pt]
  \midrule[0.5pt]
  CPU Hours & MNIS & HSCS & AIS & ACS & LRTA & ASDK & OPT. & \ours \\ \midrule[1pt]
  6T-SRAM & 4.8 & 2.0 & 4.6 & 4.6 & 3.1 & 3.2 & 2.0   &  \textbf{0.8} \\
  OTA & 50.5 & 43.0 & 26.8 & 26.5 & 24.3 & 25.0 & 10.0 &  \textbf{3.5} \\
  6-bit Array & 270.9 & 283.2 &95.9 &84.3 &73.9& 75.7 &32.4 &\textbf{9.8} \\ 
  \midrule[0.5pt]
  \bottomrule[1.5pt]
  \end{tabular}
  \end{adjustbox}
  \vspace{-0.2in}
  \end{table}


\subsection{Computational Time Study}

\begin{figure}[t]
  \centering
  \includegraphics[width=1\linewidth]{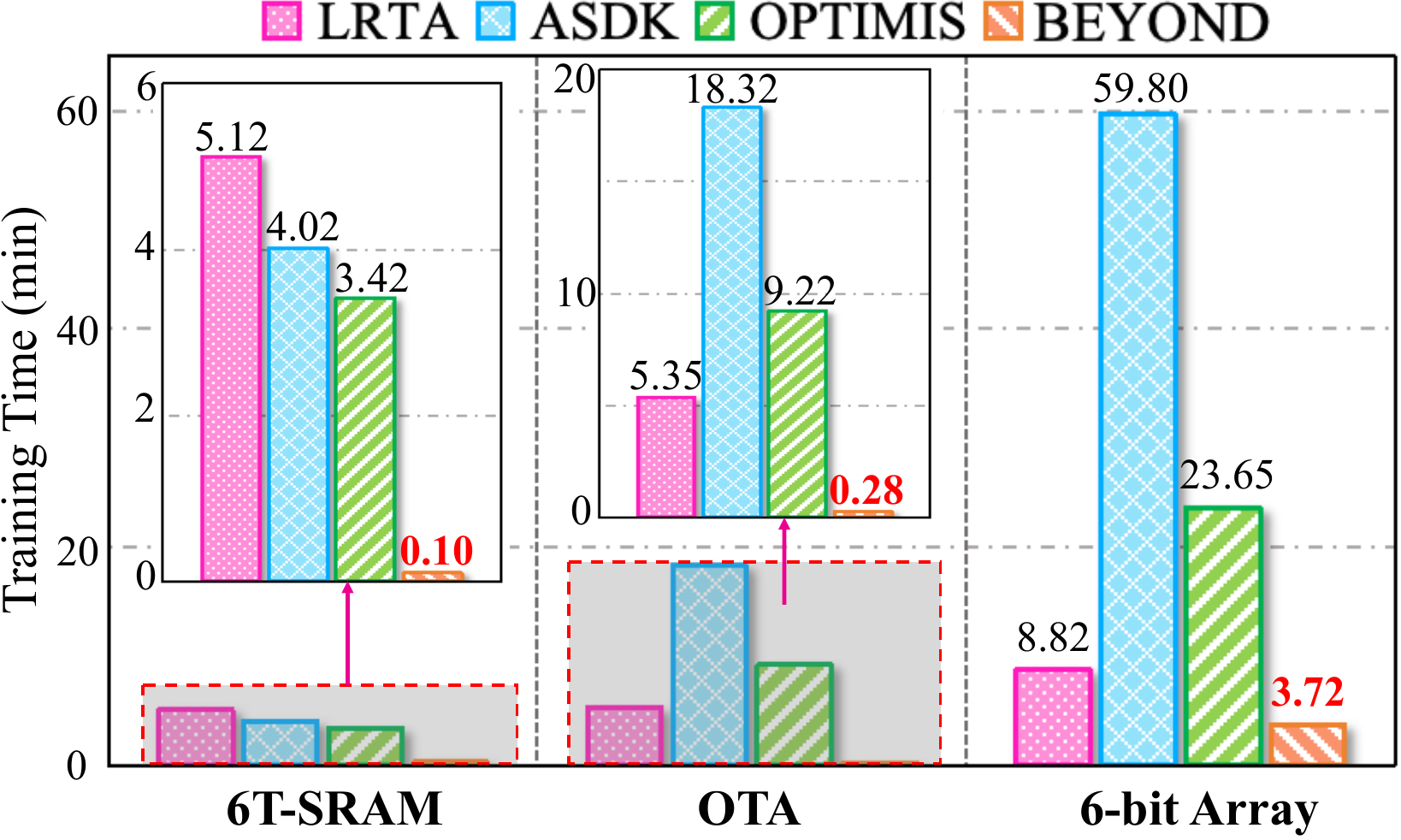}
  \vspace*{-0.25in}
  \caption{Model training time on three benchmark circuits}
  \label{fig:train}
\end{figure}

{
We demonstrate the computational time for the aforementioned yield estimation experiments in this section to highlight the efficiency.
\Tabref{tab:com_time} presents the average computational time for ten random seed runs for each method. Clearly, \ours demonstrates a leading edge in computational efficiency, showing a 4.64$\times$, 8.39$\times$, and 13.34$\times$ speedup on average for the 6T-SRAM, OTA, and 6-bit 6T-SRAM array circuits, respectively.

Furthermore, we conduct comparative experiments on the training time (find the optimal parameters) between \ours and other surrogate-based methods (LRTA, ASDK, OPTIMIS). As depicted in \Figref{fig:train}, \ours achieves a 41.87$\times$, 39.15$\times$, and 8.27$\times$ speedup on average for the 6T-SRAM, OTA, and 6-bit 6T-SRAM array circuits, respectively.

}

\subsection{Ablation Study}

\begin{figure}[h]
  \centering
  \includegraphics[width=1.03\linewidth]{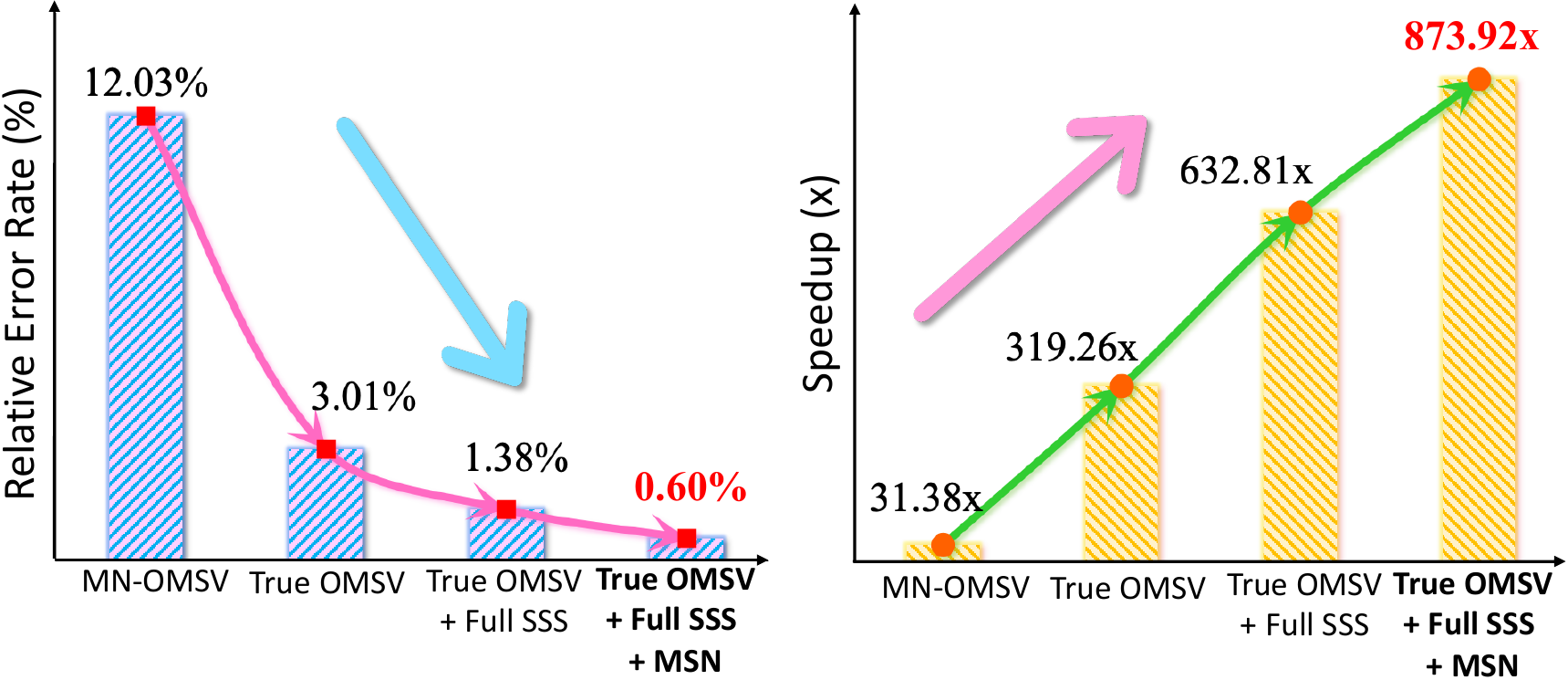}
  \vspace*{-0.25in}
  \caption{Ablation study of each component}
  \label{fig:ablation}
  \vspace*{-0.15in}
\end{figure}

To assess the contribution of each component, \ie True OMSV, Full SSS, and the mixture of skew normal distributions, we conduct an ablation study on 6-bit 6T-SRAM array (the most challenging one among our testing examples) by incrementally integrating components into the basic MN-OMSV method.
The experimental results are presented in \Tabref{tab:ablation}, which reveal consistent improvement with the incorporation of each component.
\Figref{fig:ablation} more vividly illustrates the trend of accuracy and efficiency.

True OMSV brings the most significant improvement in both accuracy and efficiency, reducing the relative error rate by 9.02\% and the number of simulations by a factor of 10.17.
Full SSS further improves the accuracy by 1.63\% and the efficiency by 2.01$\times$, whereas skew normal distribution brings a marginal improvement of 0.78\% in accuracy and 1.25$\times$ in efficiency.
These results substantiate the superiority of \ours (True OMSV+Full SSS+MSN) in yield estimation.

\begin{table}[h]
  \centering
  \caption{Ablation Study of \ours on 6-bit Array}
  \label{tab:ablation}
    \vspace{-0.15in}
    \begin{adjustbox}{width=1.03\columnwidth,center}
    \renewcommand\arraystretch{1.4}
    \setlength{\tabcolsep}{2.5pt}
  \begin{tabular}{l|cccc}
  \toprule[1.5pt]
  \midrule[0.5pt]
  \multicolumn{1}{c|}{Model} & Fail. Rate & Rel. Err. & \# Sim & Speedup \\ 
  \midrule[1pt]
  MC & 5.62e-5 & -  & 1417500 & 1$\times$ \\
  MN-OMSV & 4.94e-5 & 12.03\%  & 45174 & 31.38$\times$ \\
  True OMSV & 5.45e-5 & 3.01\% & 4440 & 319.26$\times$ \\
  True OMSV+Full SSS & 5.54e-5 & 1.38\% & 2240 & 632.81$\times$ \\
  True OMSV+Full SSS+MSN & \textbf{5.59e-5} & \textbf{0.60\%} & \textbf{1622} & \textbf{873.92$\times$}\\ 
  \midrule[0.5pt]
  \bottomrule[1.5pt]
  \end{tabular}
  \end{adjustbox}
  \vspace{-0.15in}
  \end{table}

\subsection{Robustness Study}
To highlight the robustness that is highly valued by the industry, we conduct an in-depth robustness study on the three benchmark circuits for all methods.
Specifically, each method is executed with the same set of ten consecutive random seeds, and the counts of incorrect estimations---where the relative error rate exceeds 30\%---is recorded for each method. The statistical outcomes are presented in \Tabref{tab:robust}.

In the 6T-SRAM experiments, OMSV-based methods generally demonstrate better stability than surrogate-based methods, with MNIS and \ours showing the highest stability. However, as circuit complexity increasing, as observed in the OTA and 6-bit Array experiments, the stability of all methods declines. Despite this, \ours continues to exhibit superior stability.

\begin{figure}[t]
  \centering
  \includegraphics[width=1\linewidth]{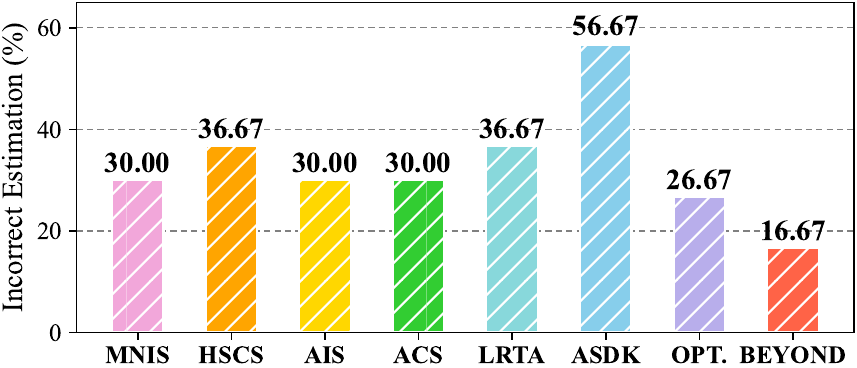}
  \vspace*{-0.25in}
  \caption{Incorrect estimation ratio in all experiments}
  \label{fig:robust}
  \vspace*{-0.17in}
\end{figure}

Based on the statistical results of incorrect estimations for the three circuits, the percentage of incorrect estimations for each method is depicted in \Figref{fig:robust}. From the percentages of incorrect estimations, it is observable that OMSV-based methods generally exhibit higher robustness compared to surrogate-based methods, with \ours being the most stable among the all methods.

\begin{table}[t]
  \centering
  \caption{The Comparison of Incorrect Estimation Counts}
  \label{tab:robust}
    \vspace{-0.15in}
    \begin{adjustbox}{width=1.\columnwidth,center}
    \renewcommand\arraystretch{1.3}
    \setlength{\tabcolsep}{2.5pt}
\begin{tabular}{l|cccccccc}
  \toprule[1.5pt]
  \midrule[0.5pt]
Circuit & MNIS & HSCS & ACS & AIS & LRTA & ASDK & OPT. & \ours \\ 
\midrule[1pt]
6T-SRAM & \textbf{1/10} & 2/10 & 2/10 & 2/10 & 3/10 & 5/10 & 2/10 & \textbf{1/10} \\
OTA & 3/10 & 3/10 & 3/10 & 3/10 & 4/10 & 7/10 & 4/10 & \textbf{2/10} \\
6-bit Array & 5/10 & 6/10 & 4/10 & 4/10 & 4/10 & 5/10 & \textbf{2/10} & \textbf{2/10} \\ 
  \midrule[0.5pt]
  \bottomrule[1.5pt]
  \end{tabular}
  \end{adjustbox}
  \vspace{-0.1in}
  \end{table} 


\section{Conclusion}

We propose \ourFramework, a rigorous analysis framework for yield estimation, which may revolutionize the traditional yield estimation paradigm.
Based on \ourFramework, we propose \ours, a novel yield estimation method.
The capacity of \ours is demonstrated by multiple modifications to the classic OMSV method in both yield estimation and optimization. 
\ours's superiority is validated by comprehensive experiments conducted on real-world circuit benchmarks, computational time studies, ablation studies and robustness studies.
With the way paved by this work, we expect more innovative methods to be developed in the future.
Dealing with high-dimensional circuits remains challenging, a common issue faced by all SOTA methods. We will further investigate the potential of multi-region sampling and dimensionality reduction strategies to address this issue.

\bibliography{OPTIMIS}

\appendix

\end{document}